\documentclass[manuscript]{aastex61}
\pdfoutput=1 
\usepackage{amsmath,amstext}
\usepackage[T1]{fontenc}
\usepackage{apjfonts}
\usepackage[figure,figure*]{hypcap}
\usepackage[version=3]{mhchem}
\usepackage{savesym}
\savesymbol{tablenum}
\usepackage{color}
\usepackage{siunitx}
\restoresymbol{SIX}{tablenum}

\shorttitle{AASTeX 6.1 Template}
\shortauthors{Lee et al.}

\begin{document}

\title{Laboratory Rotational Spectra of Silyl Isocyanide}

\author{K. L. K. Lee}
\affiliation{Harvard-Smithsonian Center for Astrophysics, 60 Garden St., Cambridge MA 02138, USA}
\email{kinlee@cfa.harvard.edu}
\author{C. A. Gottlieb}
\affiliation{Harvard-Smithsonian Center for Astrophysics, 60 Garden St., Cambridge MA 02138, USA}
\email{cgottlieb@cfa.harvard.edu}
\author{M. C. McCarthy}
\affiliation{Harvard-Smithsonian Center for Astrophysics, 60 Garden St., Cambridge MA 02138, USA}
\email{mccarthy@cfa.harvard.edu}

\correspondingauthor{K. L. K. Lee}

\begin{abstract}

The rotational spectrum of silyl isocyanide (SiH$_3$NC), an isomer of the well studied silyl cyanide (SiH$_3$CN), has been detected in the laboratory in a supersonic molecular beam, and the identification was confirmed by observations of the corresponding rotational transitions in the rare isotopic species SiH$_3$$^{15}$NC and SiH$_3$N$^{13}$C. Spectroscopic constants derived from 19 transitions between $11 - 35$~GHz in the three lowest harmonically related rotational transitions in the $K = 0 ~{\rm{and}}~1$ ladders of the normal isotopic species including the nitrogen nuclear quadrupole hyperfine constant, allow the principal astronomical transitions of SiH$_3$NC to  be calculated to an uncertainty of about 4~km~s$^{-1}$ in equivalent radial velocity, or within the FWHM of narrow spectral features in the inner region of IRC+10216 near 200~GHz. The concentration of SiH$_3$NC in our molecular beam is three times less than SiH$_3$CN, or about the same as the corresponding ratio of the isomeric pair SiNC and SiCN produced under similar conditions. Silyl isocyanide is an excellent candidate for astronomical detection, because the spectroscopic and chemical properties are very similar to SiH$_3$CN which was recently identified in the circumstellar envelope of IRC+10216  by \citet{cernicharo_discovery_2017} and of SiNC and SiCN in the same source.

\end{abstract}


\section{Introduction}
A long standing question in molecular astronomy is how silicon carbide (SiC) dust is formed in carbon-rich asymptotic giant branch (AGB) stars \citep{cherchneff_inner_2012,yasuda_formation_2012}. Following its detection, the broad unresolved absorption band at $11.3~\micron$ attributed to solid SiC \citep{hackwell_long_1972,treffers_high-resolution_1974} has become a principal spectroscopic feature for studying the evolution of dust shells in carbon stars \citep[see][and references therein]{speck_effect_2005}. Interferometric measurements of the $11.3~\micron$ band determined that the ``dust forming region'' is about five stellar radii ($5R_{\star}$ or $0.150\arcsec$) from the central star in the prototypical source IRC+10216 \citep{monnier_mid-infrared_2000}.

Radio astronomers extended the initial work on SiC in the IR by observing seven organosilanes in IRC+10216. Beginning in 1984, four silicon carbides SiC, SiCC, c-SiC$_3$, and SiC$_4$ \citep[for a summary see][and references therein]{mccarthy_silicon_2003}; the cyanide/isocyanide pair SiCN and SiNC; and the disilicon carbide SiCSi were identified in the radio band with single antennas. Most organosilanes were shown to be present in the outer molecular envelope at a radius of about $15\arcsec$ from the central star. But owing to the development in the past few years of ground based millimeter-wave interferometers with increased angular resolution and spectral bandwidths, it is now feasible to study the inner dust forming region of IRC+10216 where much less is known about the chemical composition than the well-studied outer molecular envelope. Observations with ALMA confirmed earlier indirect evidence that SiCC is also present in the inner region;\footnote{\footnotesize\citet{prieto_si_bearing_2015} determined that there is SiCC emission centered on the star with a size of about $3\arcsec \times 5\arcsec$, in addition to the SiCC gas in the outer molecular envelope at a radius of about $15\arcsec$ from the star.} two diatomic silicon bearing species are present in the inner envelope \citep[SiS and SiO:][]{bieging_submillimeter-and_2000,bieging_high_2001,prieto_si_bearing_2015}; and there is indirect evidence from single antenna observations that SiCSi is also present in the warmer inner region as well as the outer envelope \citep{cernicharo_discovery_2015}. 

To date spectral line observations in the radio band and infrared and supporting chemical models have failed to establish the phyiscochemical mechanism(s) of formation of solid silicon carbide in the dust formation zone, because most chemical reactions involving small silicon bearing molecules have not been measured in the laboratory \citep[see][]{cherchneff_inner_2012}, and the key silicon bearing reactive intermediates have not yet been identified. The recent discovery of the organosilane CH$_3$SiH$_3$ in IRC+10216 by \citet{cernicharo_discovery_2017} and confirmation of the earlier tentative detection of silyl cyande SiH$_3$CN by \citet{agundez_new_2014} has introduced a new level of chemical complexity in the inner region of this well-studied carbon rich AGB star. \citet{cernicharo_discovery_2017} estimated that the CH$_3$SiH$_3$ emission extends from $40 R_{\star}$ to at or near the outer envelope at $600 R_{\star}$, and a similar extent was found for SiH$_3$CN. Detection of CH$_3$SiH$_3$ and SiH$_3$CN suggest that other silicon bearing species of comparable size, such as SiH$_3$NC (the isocyanide isomer of SiH$_3$CN), might be within reach in the laboratory and in IRC+101216.

In this paper we report the laboratory measurements of the rotational spectrum of SiH$_3$NC (silyl isocyanide).
While there have been several studies dedicated to the rotational spectrum of SiH$_3$CN, the rotational spectrum of SiH$_3$NC had not been measured prior to this work although it had been observed by low resolution IR matrix spectroscopy \cite{maier_reaction_1998}. Our laboratory measurements of SiH$_3$NC were guided by \textit{ab initio} calculations which provided initial estimates of the rotational and hyperfine constants (Section~2), and the identification of the new species was confirmed by isotopic substitution. With our measured spectroscopic constants of SiH$_3$NC in hand, the rotational transitions of principal interest to radio astronomers can be calculated with sufficient accuracy to allow a deep search in IRC+10216 in the millimeter band. SiH$_3$NC is an excellent candidate for astronomical detection because the dipole moment and rotational partition function of SiH$_3$NC and SiH$_3$CN are very similar, and prior observations of SiCN and SiNC in IRC+10216 and the same laboratory source in which SiH$_3$NC is observed.

\section{Ab initio calculations}
\textit{Ab initio} calculations were performed with a local version of CFOUR \citep{stanton_j_f_cfour_2017}. Geometry optimizations of \ce{SiH3CN} and \ce{SiH3NC} were conducted with the coupled-cluster method with single, double, and perturbative triple excitations, combined with Dunning's correlation-consistent basis sets \citep{Dunning:1989bx} -- an approach that has been shown to provide accurate equilibrium geometries for small molecules \citep{Bak:2001jk}. All electrons are correlated in our calculations, although they utilize basis sets without core-valence functions (cc-pVXZ, where X=D,T,Q) and with the weighted core-valence basis functions of \cite{Balabanov:2005hma} (cc-pwCVXZ). Harmonic frequency analysis was carried out on the optimized geometries to confirm that they are minimum energy structures. The resulting structures calculated with quadruple-$\zeta$ (VQZ) quality basis sets are shown in Figure \ref{fig:geometry}.



The minimum energy structures of both isomers are closed-shell, prolate symmetric tops with $C_{3\mathrm{V}}$ symmetry. By comparing the quantum chemical calculations on \ce{SiH3CN} with experimental spectroscopic constants, we determined an expected accuracy for subsequent predictions of \ce{SiH3NC}. These results can be found in the Supporting Information (Table \ref{tab:sih3cn}). The inclusion of additional basis shells and core-valence functions act to increase the rotational constants $B_e$. The best agreement (0.03\%) with experiment was achieved at the CCSD(T)/cc-pwCVQZ level. Without the weighted core-valence functions (i.e. cc-pVQZ basis), the disparity between $B_e$ and $B_0$ increases by an order of magnitude. Comparisons of the \textit{ab initio} dipole and nitrogen quadrupole moment with the values determined by \citet{priem_analysis_1998} suggest that a good agreement is reached with quadrupole-$\zeta$ quality basis, regardless of core-valence functions.

The equilibrium properties of \ce{SiH3NC} calculated with different basis sets are given in Table \ref{tab:abinitio}. The rotational constants from previous calculations \citep{zanchini_silylcyanides_2000,wang_theoretical_2004} are in qualitative agreement with our CCSD(T) results, with the literature $B_e$ values within ${\sim}$1\% of our results. While the dipole moment is relatively insensitive to basis set size, the \ce{^{14}N} quadrupole coupling constant $eQq$ changes by nearly an order of magnitude between double- and triple-zeta quality bases. The change in $eQq$ between triple- and quadruple-zeta basis is considerably smaller, although it does not appear to have converged to the complete basis set limit. The inclusion of core-valence basis functions does not have a large effect on the multipole moments. Our calculations predict that the dipole moment of \ce{SiH3NC} (Table \ref{tab:abinitio}) is comparable to that of \ce{SiH3CN}, which is known to be large from Stark measurements \citep[\SI{3.44}{D};][]{priem_analysis_1998}. With the nitrogen atom located closer to the center-of-mass in \ce{SiH3NC}, the quadrupole constant ($eQq$(\ce{N})=\SI{-0.91}{\mega \hertz}) is reduced by a factor of five relative to when the nitrogen atom is at the terminal position (\ce{SiH3CN}, $eQq(\ce{N})=$\SI{-5.0440}{\mega \hertz}). Thus, we expect that the rotational spectrum of \ce{SiH3NC} should be very similar to that of \ce{SiH3CN} -- an symmetric top with $a$-type transitions but with more compact hyperfine structure.

\newpage
\section{Laboratory Measurements}

The Fourier-transform microwave (FTMW) cavity spectrometer used in this work is described in previous publications \citep{mccarthy_microwave_2000}. \ce{SiH3NC} and \ce{SiH3CN} were synthesized \textit{in situ} using a mixture of \ce{SiH4} (${\sim}0.1\%$) and \ce{CH3CN} (${\sim}0.1\%$) heavily diluted in \ce{Ne}. For detection of the rare isotopic species, either \ce{CH3{} ^{13}CN} or \ce{CH3C{}^{15}N} were used in place of \ce{CH3CN}. The gas mixture was introduced into the spectrometer via a pulsed valve with an electrical discharge nozzle \citep{mccarthy_microwave_2000}. A high voltage discharge (\SI{1.4}{\kilo \volt}) initiates a series of chemical reactions by fragmenting the precursor gases. A sufficient number of collisions occur prior to the adiabatic expansion to efficiently produce a mixture of new stable and transient species. At the orifice of the discharge nozzle, the products are rapidly cooled by adiabatic expansion in the confocal Fabry-Perot cavity, achieving a rotational temperature of ${\sim}$\SI{2}{\kelvin} near the center of the cavity.

On the basis of our new \textit{ab initio} calculations, a search for the fundamental rotational ($J_K = 1_0 - 0_0$) transition of \ce{SiH3NC} was centered near \SI{11380}{\mega \hertz}. Three unidentified, closely spaced (${\sim}$\SI{300}{\kilo \hertz}) features were found at approximately \SI{11387}{\mega \hertz} (Figure \ref{fig:fundamental}). The intensity and lineshape of these features were unaffected by a small external permanent magnet, implying that the species has a closed-shell ground state. The lines required the presence of both \ce{SiH4} and \ce{CH3CN}, and were not observed when the electrical discharge was absent. Thus, the unknown species is almost certainly a closed-shell, silicon containing molecule with closely-spaced splittings characteristic of nitrogen quadrupole hyperfine structure. Subsequent searches revealed features at integer multiples of \SI{11386}{\giga \hertz}; i.e. at \SI{22774}{} and \SI{34161}{\mega \hertz}, as expected for \ce{SiH3NC}. In all 19 transitions, with well-resolved hyperfine and $K=0$ and $1$ (albeit weak) structure, were ultimately observed. The $K=1$ transitions were approximately three times weaker than the corresponding $K=0$ lines, consistent with what is expected at a rotational temperature of 2\,K --- for this reason, no attempt was made to search for the $J_K=3_2 - 2_2$ transition near \SI{34}{\giga \hertz}. A standard symmetric top Hamiltonian with one hyperfine nucleus was used to analyze the observed frequencies. The fitting was performed with the SPFIT suite of programs \citep{pickett_fitting_1991}. The assignments and model differences are summarized in Table \ref{tab:normalfreq} and the derived spectroscopic constants in Table \ref{tab:constants}.

To remove any possible ambiguity that \ce{SiH3NC} is the carrier of the observed lines, experiments with isotopically labelled \ce{CH3CN} were subsequently performed. The optimized CCSD(T)/cc-pwCVQZ \ce{SiH3NC} structure was used to predict the rotational constants $B_e$, where the \ce{^{12}C} and \ce{^{14}N} atoms were replaced by their rare isotopes. An empirical scaling factor, determined by the ratio of the \textit{ab initio} $B_e$ and the experimental $B_0$ for the normal species, was applied to the \textit{ab initio} $B_e$ to predict the frequencies for these isotopologues. Because \ce{^{15}N} lacks a nuclear quadrupole moment ($I = \frac{1}{2}$), the hyperfine structure collapses for \ce{SiH3{^{15}}NC} and only the symmetric top transitions should be observed, while the spectrum of the \ce{^{13}C} species should be nearly identical to that of \ce{SiH3CN}, i.e. with well-resolved \ce{^{14}N} hyperfine structure. The fundamental transitions of both isotopic species were detected within \SI{0.1}{\mega \hertz} of those predicted from the scaled rotational constants, and the remaining lines at higher frequency were quickly found, resulting in a total of 16 and 5 transitions for the \ce{^{13}C} and \ce{^{15}N} species (Tables \ref{tab:c13freq} and \ref{tab:n15freq}). As with the normal species, a symmetric top Hamiltonian with or without hyperfine interactions reproduced the measured frequencies. The derived spectroscopic constants are summarized in Table \ref{tab:constants}. A search was also conducted for the two rare silicon species (\ce{^{29}Si}, 4.7\% and \ce{^{30}Si}, 3\%), although the sensitivity was not sufficient to detect either owing to their low fractional abundance.



We also estimated the relative abundances of \ce{SiH3CN} and \ce{SiH3NC} under similar experimental conditions by using the calculated CCSD(T)/cc-pwCVQZ dipole moment of \ce{SiH3NC} and the experimental value for \ce{SiH3CN} \citep[\SI{3.44}{D};][]{priem_analysis_1998}, and on the assumption the rotational temperature (\SI{2}{\kelvin}) is the same for both species. By comparing the difference in relative line intensities of the same transition ($J_K=1_0 - 0_0$, $F= 2 - 1$), we estimate that \ce{SiH3CN} is approximately three times more abundant than \ce{SiH3NC} in our discharge source.

\section{Discussion}

The laboratory measurements and calculations provide overwhelming evidence that SiH$_3$NC is the carrier of the lines shown in Figure \ref{fig:fundamental}. The measured frequency of the triplet of lines in Figure \ref{fig:fundamental} is within 0.1\% of that calculated with the equilibrium rotational constants $(B_e + C_e)$ at the CCSD(T)/cc-pwCVQZ level of theory (Table \ref{tab:abinitio}). The relative intensities and frequency separations of the closely spaced triplet is characteristic of $^{14}$N quadrupole hyperfine structure in the lowest rotational transition, and is the only triplet pattern observed in the {200}~MHz wide search range. Assays confirmed that the carrier: (1) contains Si and N, (2) is a closed-shell molecule, and (3) is only observed when there is an electrical discharge through the stable precursor gases CH$_3$CN and SiH$_4$. The observed $eQq({\rm{N}})$ for {SiH$_3$NC} is much smaller than that of {SiH$_3$CN}, as expected when the nitrogen atom is near the center-of-mass rather than at the terminal position. The theoretical value of $eQq({\rm{N}})$ for {SiH$_3$NC} agrees to better than 5\% with the measured constant. The observed $K$ rotational structure is that of a symmetric top with $C_\mathrm{3v}$ symmetry, and the frequency separation of the $K=0$ and~1 lines is comparable to that of {SiH$_3$CN}. Isotopic substitution provides the final piece of evidence in support of the identification: the SiH$_3$N$^{13}$C and SiH$_3^{15}$NC isotopic species were synthesized when $^{13}$C or $^{15}$N were substituted for either $^{12}$C or $^{14}$N in the {CH$_3$CN} precursor, and the rotational lines of the rare isotopic species were at the expected isotopic shifts in frequency.

Cyanide ({--CN}) and isocyanide ({--NC}) isomer pairs have received a great deal of attention in molecular astronomy, largely owing to the anomalously high abundance of HNC in cold dark clouds and its formation via ion-molecule reactions involving {HCNH$^+$} \citep{loison_interstellar_2014}. In the outer molecular envelope of IRC+10216 the abundances of the {--CN} isomers of the cyanopolyynes {HCN}, {HC$_3$N}, and {HC$_5$N} are $100 - 1000$ times higher than the corresponding {--NC} isomers \citep{bieging_hcn_1988,guelin_detection_2004,agundez_new_2014}. Perhaps more relevant to the work here are the metal bearing\footnote{The term ``metal bearing'' used here refers to a {\it gaseous} molecule in an evolved star which contains Al, Ti, Fe, Si, or Mg. We prefer ``metal bearing'', because in terrestrial chemistry ``refractory'' refers to a {\it solid} whose melting point (fusion temperature) is $\ge1850$~K [e.g., solid silicon carbide (SiC) or carborundum whose melting~point is 3100~K].} cyanides/isocyanide pairs {MgCN}/{MgNC} and {SiCN}/{SiNC} which are among the more abundant metal bearing molecules in IRC+10216. Both pairs have been observed in the outer envelope of IRC+10216 \citep{ziurys_l.m._detection_1995,guelin_detection_2004}, but have significantly different abundances despite comparable heats of formation ($\Delta H$). The {MgNC}/{MgCN} ratio of ${\sim}$20 was rationalized by \cite{ziurys_l.m._detection_1995} on the basis of  $\Delta H$ of {MgNC} which is lower than {MgCN} by 950~K \citep{senent_cyanide/isocyanide_2012}. However, the same argument does not  apply to {SiCN} and {SiNC}, because the column densities of the two silicon bearing species are comparable, despite a relative $\Delta H$ that is similar to {MgNC}/{MgCN}. Only one other cyanide/isocyanide isomeric pair (HCN/HNC) has been observed in the inner envelope of IRC+10216. Close to the photosphere, the abundance of HNC is fairly high and the {HCN}/{HNC} ratio is about 10, but the ratio increases rapidly with distance from the star \citep{cernicharo_unveiling_2013}. Lacking possible analogs in the inner envelope like that of {SiCN}/{SiNC}, estimation of the SiH$_3$CN/SiH$_3$NC ratio in the inner envelope awaits detection of SiH$_3$NC.

There are no apparent obstacles for detection of {SiH$_3$NC} in IRC+10216: the dipole moments, rotational partition functions, and spectroscopic constants of {SiH$_3$NC} and {SiH$_3$CN} are similar. We did not extend the measurements of SiH$_3$NC to higher frequency, instead the effect of neglecting high order centrifugal distortion terms was estimated by referring to {SiH$_3$CN} which has been measured both in the centimeter and millimeter bands \citet{priem_analysis_1998}. We fit the four spectroscopic constants in the truncated Hamiltonian used to analyze SiH$_3$NC (see Table~\ref{tab:normalfreq}) to the centimeter-wave measurements of {SiH$_3$CN} alone, and compared the frequencies of SiH$_3$CN  calculated with this Hamiltonian to the measured ones in the millimeter band. The estimated uncertainty in the calculated frequencies of {SiH$_3$NC} at 200~GHz --- i.e.,  near the highest frequency where {SiH$_3$CN} was observed in IRC+10216 \citep{cernicharo_discovery_2017} --- is about 4~km~s$^{-1}$, or a fraction the FWHM of the features observed in the inner region of IRC+10216 with ALMA \citep{cernicharo_unveiling_2013}.

Because silicon carbide films are synthesized in the laboratory from {CH$_3$SiH$_3$} and other substituted methyl silanes [(CH$_3$)$_n$SiH$_{(4-n)}$], it is worth considering whether solid SiC is formed by the same chemical mechanism in the laboratory and the inner envelope of a carbon-rich AGB star. Silicon carbide films are synthesized in the laboratory by the process known as hot wire chemical vapor deposition \citep[HWCVD:][]{zaharias_detecting_2006,shi_role_2017}. Three main processes occur in HWCVD: (1) formation of initial reactive species by activation of the precursor gas (CH$_3$SiH$_3$) on a metal filament surface heated to $\lesssim 2000$~K; (2) chemical reactions in the gas phase between the reactive species and the abundant precursor {CH$_3$SiH$_3$}; and (3) reactions on the substrate surface at temperatures near $600 - 700$~K. Although the temperatures are comparable in the inner envelope and HWCVD, the gas density in HWCVD is $10^3 - 10^4$ times higher than in the photosphere of the central star in IRC+10216 and $10^3$ times higher still than the dust forming region \citep{agundez_molecular_2012}. Another source of uncertainty in the analogy with HWCVD is whether the gas phase reactions that occur in the inner envelope of the carbon star are the same as those in the laboratory reactor. Nevertheless it is striking that the gas phase precursor (CH$_3$SiH$_3$), synthesized product (solid {SiC}), and temperatures are roughly the same in HWCVD and the inner envelope of IRC+10216.

Our laboratory detection of {SiH$_3$NC} suggests that fragments of {CH$_3$SiH$_3$}, {SiH$_3$CN}, and {SiH$_3$NC} might be key reactive intermediates in the formation of solid {SiC} dust in circumstellar envelopes. The rotational spectra of most of these organosilane fragments have not been measured in the laboratory, although some have been observed trapped in cold matrices by low resolution IR spectroscopy. \citet{kaiser_laboratory_2005} and \citet{bennett_infrared_2005} considered seven possible derivatives of CH$_3$SiH$_3$ with the elemental composition [{C},{H$_x$,{Si} where $x=3,4,5$], and provided calculations of their dipole moments and rotational constants. To date the rotational spectrum of only one fragment ({CH$_2$SiH$_2$}) has been measured in the laboratory \citep{bailleux_equilibrium_1997}, but the lowest rotational transition ($1_{0,1} - 0_{0,0}$) of the six remaining candidates are predicted to lie within range of our sensitive FTMW spectrometer used to observe SiH$_3$NC. Although the dipole moments of the [{C},{H$_x$},{Si}] fragments are modest \citep[$\mu \leq {1}$~{D}][]{kaiser_laboratory_2005}, they are comparable to the recently detected parent molecule CH$_3$SiH$_3$ (0.7~{D}) and the dipole moments of fragments of {SiH$_x$CN} and {SiH$_x$NC} ($x = 1, 2, 3$) are expected to be much larger. Another intriguing candidate for detection is {HNCSi}, a small silicon bearing species observed in abundance in a solid {Ar} matrix \citep{maier_reaction_1998}. The rotational spectrum of {HNCSi} has not yet been measured, but accurate quantum chemical predictions of its structure and spectroscopic properties are available \citep{thorwirth_coupled-cluster_2009}, and a laboratory search would appear straightforward.

\software{SPFIT (Pickett 1991), CFOUR (Stanton et al. 2017)}

\bibliographystyle{yahapj}
\bibliography{references}

\begin{figure}
	\begin{center}
      \includegraphics[width=5cm]{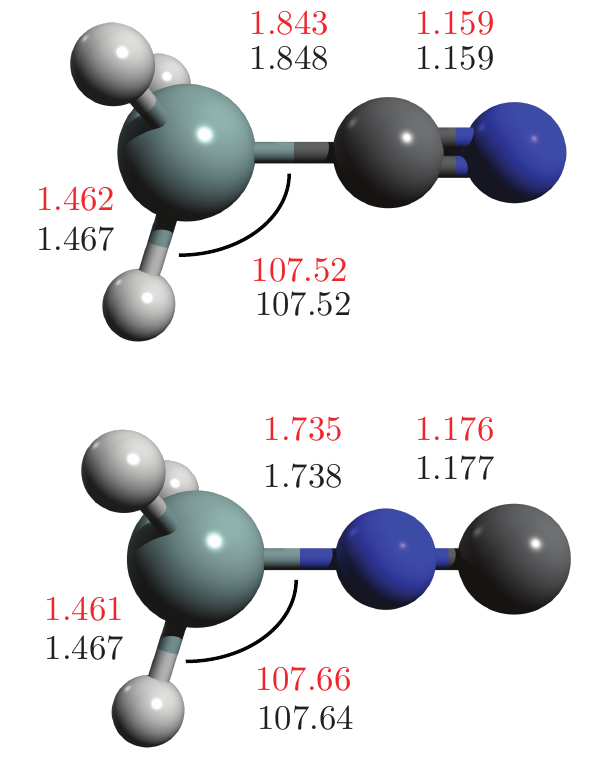}
      \caption{\textit{Ab initio} equilibrium molecular structures for \ce{SiH3CN} (top) and \ce{SiH3NC} (bottom). Values in {\color{red}red} were obtained with a cc-pVQZ basis, while the calculated values in black include core-valence basis functions (cc-pwCVQZ). Bond lengths are given in \AA, and angles in degrees. See SI for analogous calculations of \ce{SiH3CN}.}
    \end{center}
    \label{fig:geometry}
\end{figure}

\begin{deluxetable}{cccccc}
  \tablecaption{Equilibrium rotational constants ($A_e$ and $B_e$), dipole moment ($\mu_a$), and \ce{^{14}N} quadrupole moment ($eQq$) of \ce{SiH3NC} calculated at the CCSD(T) level with six basis sets.\label{tab:abinitio}}

  \tablehead{\colhead{Basis set} & \colhead{$A_e$} & \colhead{$B_e$} & \colhead{$\mu_a$} & \colhead{$eQq$} \\
\colhead{} & \colhead{(\si{\mega \hertz})} & \colhead{(\si{\mega \hertz})} & (D) & \colhead{(\si{\mega \hertz})}}

  \startdata
    cc-pVDZ & \SI{83714.42}{} & \SI{5460.07}{} & -3.22 & -0.09 \\
    cc-pVTZ & \SI{85004.89}{} & \SI{5639.51}{} & -3.17 & -0.91 \\
    cc-pVQZ & \SI{86263.47}{} & \SI{5705.69}{} & -3.15 & -1.00 \\
     &  &  &  &  &  \\
    cc-pwCVDZ & \SI{84459.22}{} & \SI{5512.48}{} & -3.15 & -0.29 \\
    cc-pwCVTZ & \SI{85631.70}{} & \SI{5674.15}{} & -3.12 & -0.87 \\
    cc-pwCVQZ\tablenotemark{a} & \SI{85575.16}{} & \SI{5688.45}{} & -3.17 & -0.91 \\
  \enddata
  \tablenotetext{a}{The basis set used to predict the \ce{SiH3NC} equilibrium rotational constants.}
\end{deluxetable}

\begin{figure}
  \begin{center}
    \includegraphics{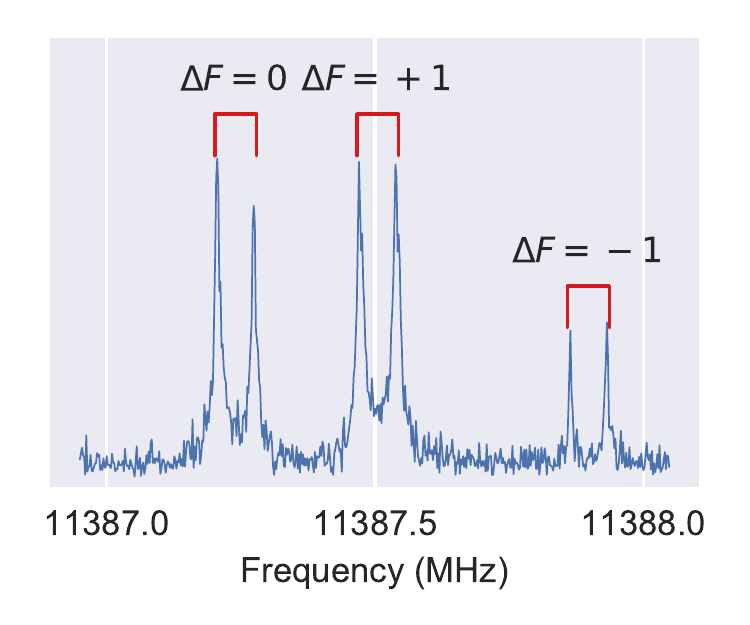}
    \caption{Sample composite spectrum of the fundamental $J_K = 1_0-0_0$ transition of \ce{SiH3NC}, showing well-resolved nitrogen quadrupole structure. Because the molecular beam expands coaxially with the Fabry-Perot cavity, each hyperfine transition is observed as a Doppler doublet; the rest frequency is the center frequency of each Doppler pair. The integration time in each frequency window was three minutes at a repetition rate of \SI{6}{\hertz}.}
  \end{center}
  \label{fig:fundamental}
\end{figure}

\begin{deluxetable}{ccclc}[h!]
\tablecaption{Laboratory rotational frequencies of \ce{SiH3NC}.\label{tab:normalfreq}}
\tablehead{\colhead{$J' - J''$} & \colhead{$K' - K''$} & \colhead{$F' - F''$} & \colhead{Frequency} & \colhead{$O - C$}\tablenotemark{a} \\ 
\colhead{} & \colhead{} & \colhead{} & \colhead{(MHz)} & \colhead{(kHz)} } 

\startdata
1 - 0 & 0 - 0 & 1 - 1 & \SI{11387.240}{} & 0 \\
 &  & 2 - 1 & \SI{11387.506}{} & 2 \\
 &  & 0 - 1 & \SI{11387.898}{} & -1 \\
2 - 1 & 1 - 1 & 2 - 1 & \SI{22774.305}{} & -3 \\
 &  & 3 - 2 & \SI{22774.577}{} & -3 \\
 & 0 - 0 & 2 - 2 & \SI{22774.614}{} & -1 \\
 &  & 1 - 0 & \SI{22774.661}{} & 2 \\
 &  & 2 - 1 & \SI{22774.873}{} & -4 \\
 &  & 3 - 2 & \SI{22774.898}{} & 2 \\
 &  & 1 - 1 & \SI{22775.319}{} & 3 \\
3 - 2 & 1 - 1 & 3 - 3 & \SI{34161.494}{} & 3 \\
 &  & 3 - 2 & \SI{34161.635}{} & 3 \\
 &  & 4 - 3 & \SI{34161.713}{}\tablenotemark{b} & 5 \\
 &  & 2 - 1 & \SI{34161.713}{}\tablenotemark{b} & 5 \\
 & 0 - 0 & 3 - 3 & \SI{34161.931}{} & 3 \\
 &  & 2 - 1 & \SI{34162.168}{} & 1 \\
 &  & 3 - 2 & \SI{34162.210}{} & 0 \\
 &  & 4 - 3 & \SI{34162.226}{} & -1 \\
 &  & 2 - 2 & \SI{34162.603}{} & 4 \\
\enddata
\tablenotetext{a}{Calculated with the spectroscopic constants from Table \ref{tab:constants}.}
\tablenotetext{b}{Blended lines.}
\end{deluxetable}

\begin{deluxetable}{ccclc}[h!]
\tablecaption{Laboratory rotational frequencies for \ce{SiH3N^{13}C}. \label{tab:c13freq}}
\tablehead{\colhead{$J' - J''$} & \colhead{$K' - K''$} & \colhead{$F' - F''$} & \colhead{Frequency} & \colhead{$O - C$}\tablenotemark{a} \\ 
\colhead{} & \colhead{} & \colhead{} & \colhead{(MHz)} & \colhead{(kHz)} } 

\startdata
1 - 0 & 0 - 0 & 1 - 1 & \SI{10950.495}{} & 0 \\
 &  & 2 - 1 & \SI{10950.762}{} & 3 \\
 &  & 0 - 1 & \SI{10951.154}{} & -1 \\
2 - 1 & 1 - 1 & 2 - 1 & \SI{21900.836}{} & -5 \\
 & 0 - 0 & 2 - 2 & \SI{21901.127}{} & 0 \\
 &  & 1 - 0 & \SI{21901.174}{} & 3 \\
 &  & 2 - 1 & \SI{21901.385}{} & -6 \\
 &  & 3 - 2 & \SI{21901.417}{} & 7 \\
 &  & 1 - 1 & \SI{21901.831}{} & 0 \\
3 - 2 & 1 - 1 & 3 - 2 & \SI{32851.441}{} & 1 \\
 &  & 4 - 3 & \SI{32851.519}{}\tablenotemark{b} & 1 \\
 &  & 2 - 1 & \SI{32851.519}{}\tablenotemark{b} & 3 \\
 & 0 - 0 & 3 - 3 & \SI{32851.702}{} & -4 \\
 &  & 2 - 1 & \SI{32851.948}{} & 2 \\
 &  & 3 - 2 & \SI{32851.984}{} & -5 \\
 &  & 4 - 3 & \SI{32852.006}{} & 6 \\
 &  & 2 - 2 & \SI{32852.383}{} & -2 \\
\enddata
\tablenotetext{a}{Calculated with the spectroscopic constants from Table \ref{tab:constants}.}
\tablenotetext{b}{Blended lines.}
\end{deluxetable}

\begin{deluxetable}{cclc}[h!]
\tablecaption{Laboratory frequencies for $\mathrm{SiH_3{^{15}N}C}$. \label{tab:n15freq}}
\tablehead{\colhead{$J' - J''$} & \colhead{$K' - K''$} & \colhead{Frequency} & \colhead{$O-C$} \tablenotemark{a} \\ 
\colhead{} & \colhead{} & \colhead{(MHz)} & \colhead{(kHz)} } 

\startdata
1 - 0 & 0 - 0 & 11321.343 & -2 \\
2 - 1 & 1 - 1 & 22642.310 & 2 \\
 & 0 - 0 & 22642.646 & -1 \\
3 - 2 & 1 - 1 & 33963.353 & -1 \\
 & 0 - 0 & 33963.864 & 1 \\
\enddata
\tablenotetext{a}{Calculated with the spectroscopic constants from Table \ref{tab:constants}.}

\end{deluxetable}

\begin{deluxetable}{ccccc}[h!]
\tablecaption{Spectroscopic constants of \ce{SiH3NC}, \ce{SiH3{^{15}}NC}, and \ce{SiH3N{^{15}}C}, derived from measurements in Tables \ref{tab:normalfreq} - \ref{tab:n15freq}. \label{tab:constants}}
\tablehead{\colhead{Parameter} & \colhead{\ce{SiH3NC}} & \colhead{\ce{SiH3N^{13}C}} & \colhead{\ce{SiH3^{15}NC}} \\ 
\colhead{} & \colhead{} & \colhead{} & \colhead{} } 

\startdata
$B$ & 5693.73300(43) & 5475.36068(69) & 5660.67605(59) \\
$D_J \times 10^3$ & 1.716(28) & 1.619(44) & 1.793(38) \\
$D_{JK}$ & 0.08736(28) & 0.08244(47) & 0.08473(39) \\
$eQq$ & -0.8765(26) & -0.8803(40) & - \\
\enddata
\tablecomments{Constants given in \si{\mega \hertz}. Values in parentheses represent $1\sigma$ uncertainty.}
\end{deluxetable}

\clearpage

\appendix
\section{\textit{Ab initio} calculations on cyanosilane}

The calculations on \ce{SiH3CN} were used to determine the expected accuracy for predicting \ce{SiH3NC}. All calculations were performed at the all-electron CCSD(T) level. The overall trend with increasing basis size and inclusion of weighted core-valence basis functions (Section 2) is increasing rotational constant $B_e$ with the number of basis functions. The excellent agreement between the experimental and \textit{ab initio} value with the quadruple-$\zeta$ quality basis is consistent with the benchmarking performed by \cite{Bak:2001jk}. The inclusion of core-valence basis functions (cc-pwCVQZ) yields the best agreement between theory and experiment with $B_e$ within 0.03\% of $B_0$. The dipole moment and \ce{^{14}N} quadrupole coupling constant is also in quantitative agreement with the experimentally determined values \citep{priem_analysis_1998}, providing confidence in our calculations of \ce{SiH3NC}.

\begin{deluxetable}{cccccc}[h!]
\tablecaption{\textit{Ab initio} equilibrium rotational constants ($A_e$, $B_e$), dipole moments, and \ce{^{14}N} quadrupole coupling constants for \ce{SiH3CN} calculated at the CCSD(T) level. \label{tab:sih3cn}}
\tablehead{\colhead{Basis set} & \colhead{$A_e$} & \colhead{$B_e$} & \colhead{$B_e - B_0$}\tablenotemark{a} & \colhead{Dipole moment} & \colhead{$eQq(\mathrm{N})$} \\ 
\colhead{} & \colhead{(MHz)} & \colhead{(MHz)} & \colhead{(MHz)} & \colhead{(D)} & \colhead{(MHz)} } 

\startdata
cc-pVDZ & \SI{83593.78}{} & \SI{4813.18}{} & -159.83 & -3.25 & -4.08 \\
cc-pVTZ & \SI{84865.07}{} & \SI{4940.02}{} & -32.99 & -3.38 & -5.16 \\
cc-pVQZ & \SI{86023.03}{} & \SI{4990.37}{} & 17.37 & -3.45 & -5.27 \\
 &  &  &  &  &  \\
cc-pwCVDZ & \SI{84256.93}{} & \SI{4847.73}{} & -125.27 & -3.23 & -4.45 \\
cc-pwCVTZ & \SI{85415.44}{} & \SI{4957.13}{} & -15.88 & -3.12 & -5.10 \\
cc-pwCVQZ\tablenotemark{b} & \SI{85368.29}{} & \SI{4971.61}{} & -1.39 & -3.46 & -5.11 \\
\enddata
\tablenotetext{a}{$B_0$ corresponds to the experimental rotational constant (\SI{4973.005714}{\mega \hertz}) determined by \cite{priem_analysis_1998}.}
\tablenotetext{b}{The basis set used to predict the \ce{SiH3NC} equilibrium rotational constants.}
\end{deluxetable}

\end{document}